\documentclass[prc]{revtex4}
\usepackage{graphicx,amsmath}

\newcommand{\gsim}{\stackrel{\scriptstyle >}{\phantom{}_{\sim}}}
\begin{document}
\title{On Cooling of Neutron Stars With Stiff Equation of State Including Hyperons}
\author{Hovik Grigorian}
\affiliation{
	Laboratory for Information Technologies, Joint Institute for Nuclear Research, RU-141980 Dubna, Russia}
\affiliation{Yerevan State University, Alek Manyukyan 1, 0025 Yerevan, Armenia}
\author{Evgeni E. Kolomeitsev}
\affiliation{Matej Bel University, Tajovskeho 40, SK-97401 Banska Bystrica, Slovakia}
\author{Konstantin A. Maslov}
\affiliation{Bogoliubov Laboratory for Theoretical Physics, Joint Institute
	for Nuclear Research, RU-141980 Dubna, Russia}
\affiliation{National Research Nuclear  University (MEPhI), Kashirskoe shosse 31, RU-115409 Moscow, Russia}
\author{Dmitry N. Voskresensky}
\affiliation{Bogoliubov Laboratory for Theoretical Physics, Joint Institute
	for Nuclear Research, RU-141980 Dubna, Russia}
\affiliation{National Research Nuclear  University (MEPhI), Kashirskoe shosse 31, RU-115409 Moscow, Russia}


\begin{abstract}
The existence of  high  mass ($\sim 2 M_{\odot}$)  pulsars  PSR  J1614-2230
and PSR J0348-0432 requires the compact star matter to be described by a stiff equation
of  state (EoS).  Presence of hyperons in neutron stars leads to a softening of the EoS
that results in a decrease of the maximum neutron-star mass below the measured values of masses for PSR  J1614-2230
and PSR J0348-0432 pulsars, if one exploits ordinary relativistic mean-field (RMF) models (hyperon puzzle). However,
within  a RMF EoS with a $\sigma$ scaled hadron effective masses and coupling constants the maximum neutron-star mass remains above $2M_{\odot}$ even when hyperons are included. Also other important constraints on the equation of state, e.g. the flow constraint from heavy-ion collisions are to be fulfilled.
We demonstrate how a satisfactory explanation of all existing observational data for the
temperature-age relation is reached within  the “nuclear medium cooling”
scenario with  a relativistic-mean-field EoS with a $\sigma$-scaled hadron effective masses and coupling constants including hyperons.
\end{abstract}

\keywords{neutron stars; equation of state; in-medium effects; hyperons; neutrino}

\maketitle

\section{Introduction}
Equation of state (EoS) of the cold hadronic matter should:
\begin{itemize}
\item satisfy experimental information on properties of dilute nuclear matter;
\item empirical constraints on global characteristics of atomic nuclei;
\item
constraints on the pressure of the nuclear mater  from the description of particle transverse and elliptic flows and the $K^+$ production in heavy-ion collisions, cf.  \cite{Danielewicz:2002pu,Lynch:2009vc};
\item
allow for the heaviest known   pulsars PSR  J1614-2230 (of mass $M = 1.928 \pm 0.017 \, M_\odot$) \cite{Fonseca:2016tux} and  PSR~J0348+0432 (of mass $M = 2.01 \pm 0.04 \, M_\odot$)~\cite{Antoniadis:2013pzd};
\item
allow for an adequate description of the compact star cooling \cite{Blaschke:2004vq}, most probably without direct Urca (DU) neutrino processes in the majority of the known pulsars detected in soft $X$ rays \cite{Klahn:2006ir};
\item
yield a mass-radius relation comparable with the empirical constraints including  recent gravitation wave LIGO-Virgo detection \cite{TheLIGOScientific:2017qsa};
\item
being extended to non-zero temperature $T$ (for $T<T_c$ where $T_c$ is the critical temperature of the deconfinement),  appropriately describe supernova explosions,  proto-neutron stars, and heavy-ion collision data, etc.
\end{itemize}
The most difficult task is to satisfy simultaneously the heavy-ion-collision flow and the maximum neutron-star mass constraints. The fulfillment of the flow constraints~\cite{Danielewicz:2002pu,Lynch:2009vc} requires a rather soft EoS of isospin-symmetric matter (ISM), whereas the EoS of the beta-equilibrium  matter (BEM) should be stiff in order to predict the maximum mass of a neutron star to be higher than the measured mass $M = 2.01 \pm 0.04 \, M_\odot$~\cite{Antoniadis:2013pzd}  of the  pulsar PSR~J0348+0432, being the heaviest among the known pulsars.

\section{Equation of state and pairing gaps}

\begin{figure}
\centering
\includegraphics[width=5.2cm]{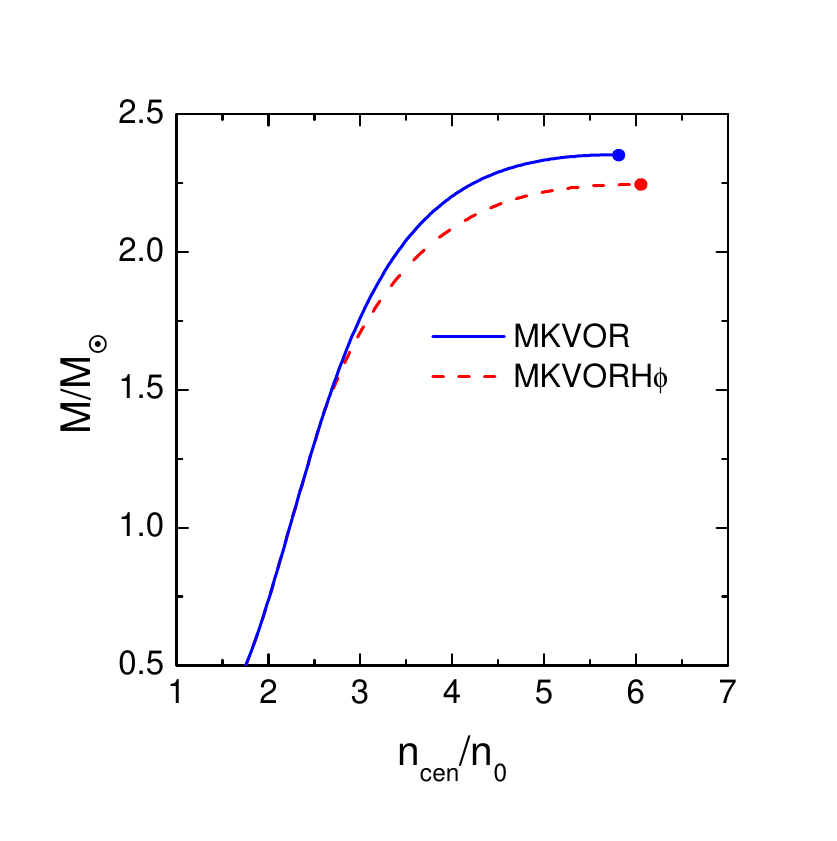}
\caption{Neutron star masses versus the central density for the MKVOR model without inclusion of hyperons and for the MKVORH$\phi$ model with included hyperons.}
\label{mkv_mass}
\end{figure}

In standard RMF models hyperons and $\Delta$-isobars  may appear in neutron-star cores already for $n\gsim (2-3) n_0$, which results in a decrease of the maximum neutron-star mass below the observed limit. The problems were named the hyperon puzzle \cite{SchaffnerBielich:2008kb,Djapo:2008au}. Within the RMF models with the $\sigma$ field-dependent hadron effective masses and coupling constants the hyperon puzzle is resolved, see \cite{Maslov:2015msa,Maslov:2015wba}. Here we use the MKVOR-based models from these works. Most of other constraints on the EoS including the flow constraints are also appropriately satisfied. In Fig~\ref{mkv_mass} we demonstrate the neutron star mass as a function of the central density for the MKVOR model without hyperons and for the MKVORH$\phi$ model with includes hyperons, cf. Fig.~20 and 25 in~\cite{Maslov:2015wba}. For MKVOR model the maximum neutron-star mass reaches $2.33 M_{\odot}$ and the DU reaction is allowed for $M>2.14 M_{\odot}$. For MKVORH$\phi$ model the maximum neutron-star mass is $2.22 M_{\odot}$. The DU reactions on $\Lambda$ hyperons $\Lambda \to p+e+\bar{\nu}$, $p+e\to \Lambda +\nu,$  become allowed for $M>1.43 M_{\odot}$. The DU reactions with participation of $\Xi^-$, $\Xi^-\to \Lambda +e+\bar{\nu}$ and $\Lambda +e\to \Xi^{-} +\bar{\nu}$ become allowed f
or $M>1.65 M_{\odot}$. However, the neutrino emissivity in these processes is not as high as for the standard DU processes on nucleons due to a smaller coupling for the hyperons.
Below we use MKVOR and MKVORH$\phi$  EoSs for calculations of the cooling history of neutron stars.

We adopt here all cooling inputs such as the neutrino
emissivities, specific heat, crust properties, etc., from our earlier works   performed on the basis of the HHJ equation of state (EoS) \cite{Blaschke:2004vq,Grigorian:2005fn,Blaschke:2011gc}, a stiffer  HDD EoS \cite{Blaschke:2013vma} and even more stiffer DD2 and DD2vex EoSs \cite{Grigorian:2016leu} for the hadronic matter. These works exploit the nuclear medium cooling scenario where the most efficient processes  are the medium modified Urca (MMU)
processes, $nn\to npe\bar{\nu}$ and $np\to ppe\bar{\nu}$, medium modified nucleon bremstrahlung (MNB) processes $nn\to nn\nu\bar{\nu}$, $np\to np\nu\bar{\nu}$, $pp\to pp\nu\bar{\nu}$, and the pair-breaking-formation  (PBF) processes $n\to n\nu\bar{\nu}$
and $p\to p\nu\bar{\nu}$. The latter processes are allowed only in supefluid matter.

\begin{figure}
\centering
\parbox{5.2cm}{\includegraphics[width=5.2cm]{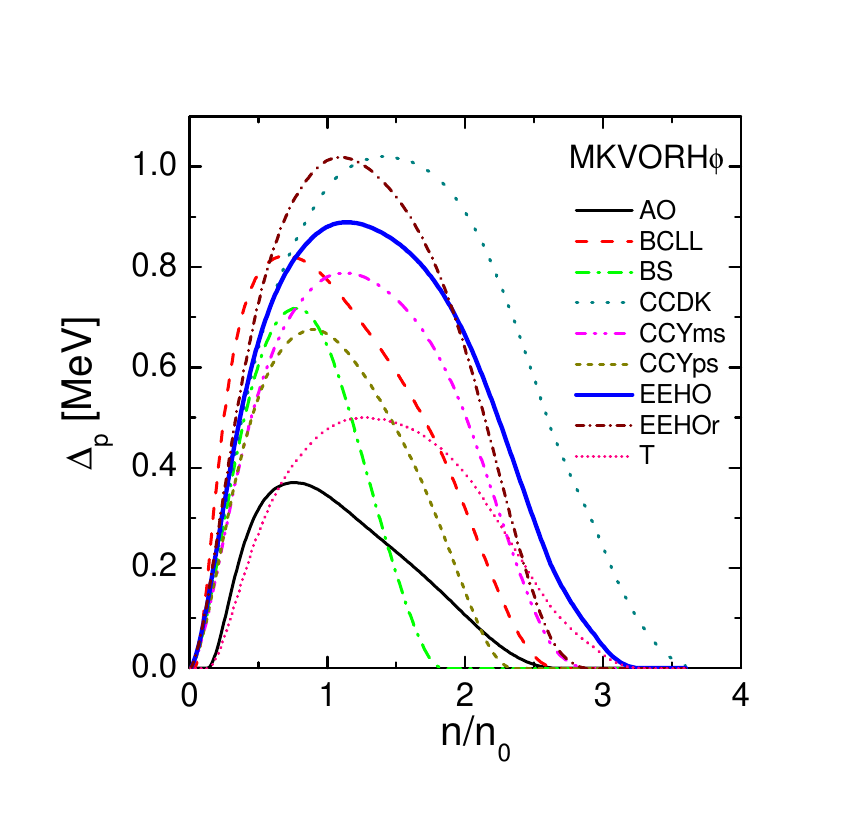}} \quad
\parbox{5.2cm}{\includegraphics[width=5.2cm]{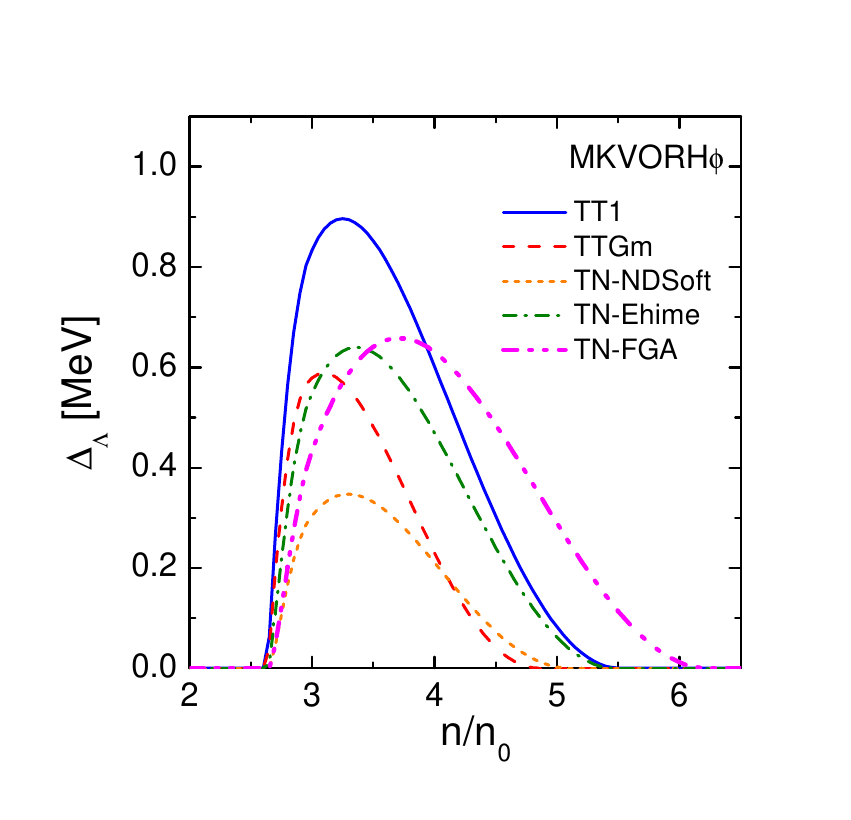}}
\caption{Pairing gaps for protons (left panel) and $\Lambda$ hyperons (right panel) as functions of baryon density for the  MKVORH$\phi$ EoS including hyperons. Proton gaps are evaluated using the same models as in~\cite{Grigorian:2016leu} and the $\Lambda$ hyperon gaps are from~\cite{TT00,TN06}.}
\label{Protongaps}
	\end{figure}

 The results are rather insensitive to the value of the $nn$ pairing gap since the $^1S_0$ neutron pairing does not spread in the interior region of the neutron star. We use the same values as we have used in our previous works. Within our scenario we continue to exploit tiny values of the $^3P_2$ $nn$ pairing gap. For calculation of the proton pairing gaps we use the same models as in  \cite{Grigorian:2016leu} but now we exploit EoS of the MKVORH$\phi$ model. The corresponding gaps are shown on the left panel of in Fig. \ref{Protongaps}.
With the increase of the density in the MKVORH$\phi$ model the $\Lambda$ hyperons are the first to appear at the density $n_c^{(\Lambda)} = 2.63 \, n_0$, and then the $\Xi^-$ hyperons appear at $n_c^{(\Xi^-)} = 2.93 \, n_0$. We take the values of the $\Lambda$ gaps from the calculations~\cite{TT00,TN06}.
The model TT1 uses the bare ND-soft model by the Nijmegen group for $\Lambda\Lambda$ interaction and model TTGm uses results of G-matrix calculations by Lanskoy and Yamamoto~\cite{LanskYam} at density $2.5 n_0$. The other 3 models include three-nucleon forces TNI6u forces for several $\Lambda\Lambda$ pairing potentials: ND-Soft, Ehime and FG-A. On the right panel we show the $\Lambda$ hyperon pairing gaps which we exploit in this work.
$\Xi^-$ are considered unpaired.

\begin{figure}
\centering
\includegraphics[height=6.5cm,clip=true]{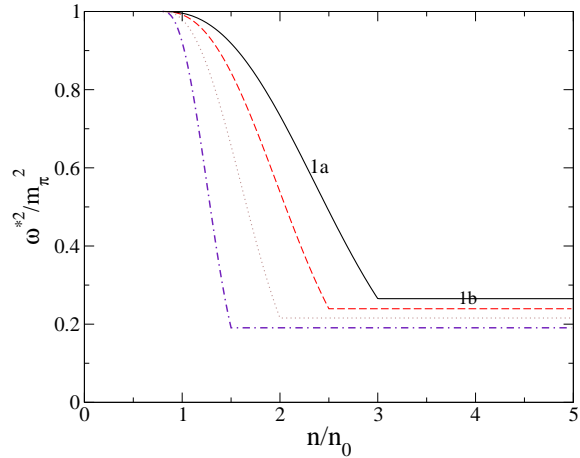}
\caption{Density dependence of the effective pion gap squared used in the given work. We assume that the pion softening effect saturates above a critical density which value we vary from $1.5$ till $3 n_0.$}
\label{piongap}
\end{figure}

 The quantity
\begin{eqnarray}
-G^{R-1}_{\pi}(\mu_\pi ,k, n)=\omega^{*2} (k)=k^2 +m_\pi^2 -\mu_\pi^2 +\mbox{Re}\Sigma_\pi (\mu_\pi ,k, n)
\end{eqnarray}
in a dense neutron-star matter (for $n\gsim n_0$) has a minimum for  $k=k_m\simeq p_{\rm F}$, where $p_{\rm F}$ is the neutron Fermi momentum. For $\pi^0$ the minimum occurs for $\mu_\pi =0$. The value $\omega^{*2} (k_m)$
has the meaning of the squared effective pion gap.  Of key importance is that we  use here the very same density dependence of the effective pion gap $\omega^{*} (n)$
as in our  previous works, e.g., see Fig. 2 of \cite{Grigorian:2016leu}.
To be specific we assume a saturation of the pion softening for $n>n_c$. We plot this pion gap in Fig. ~\ref{piongap}.

\section{Results}

\begin{figure}
\centerline{
\includegraphics[height=7.1cm,clip=true]{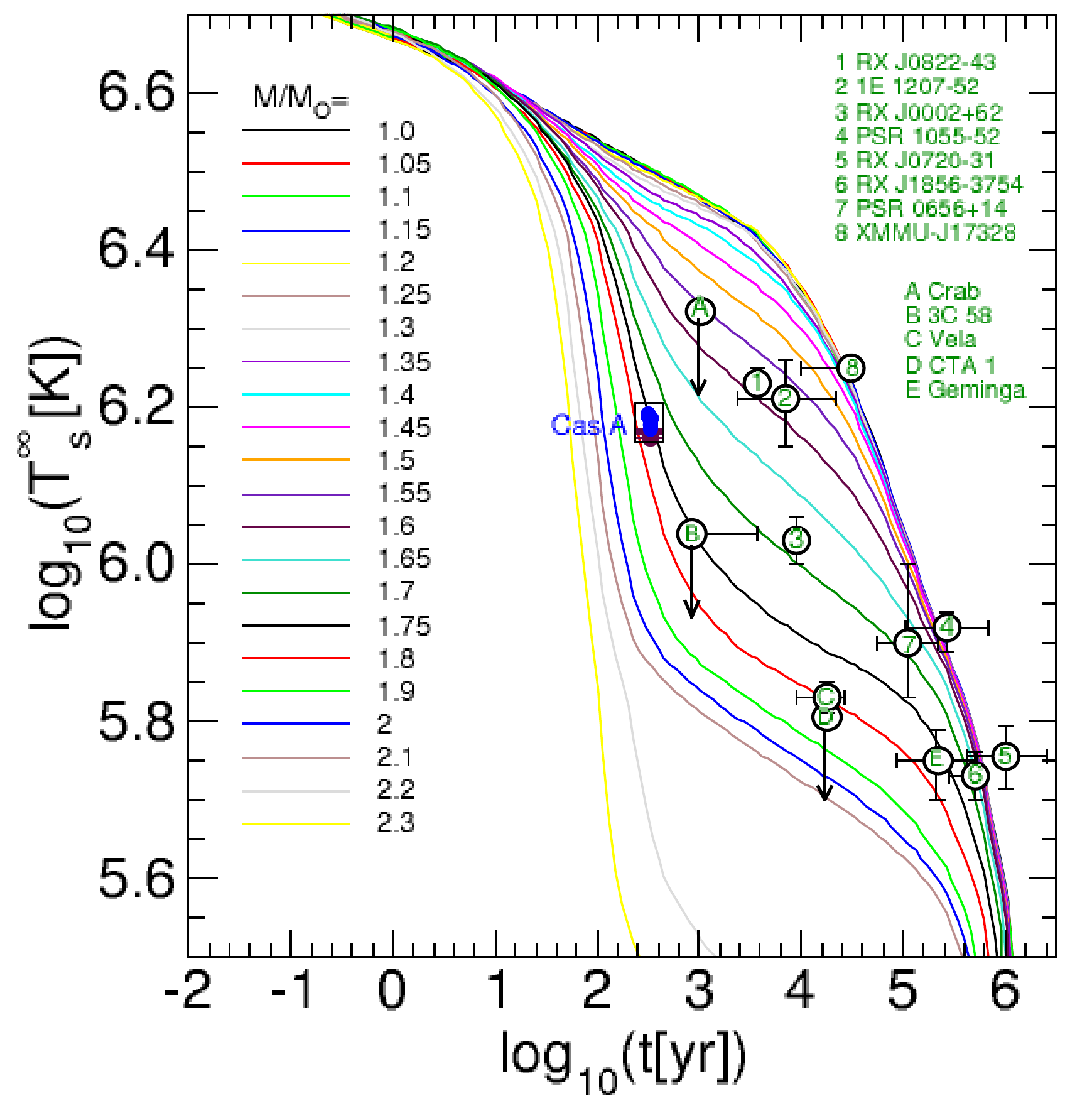}\quad
\includegraphics[height=7.1cm,clip=true]{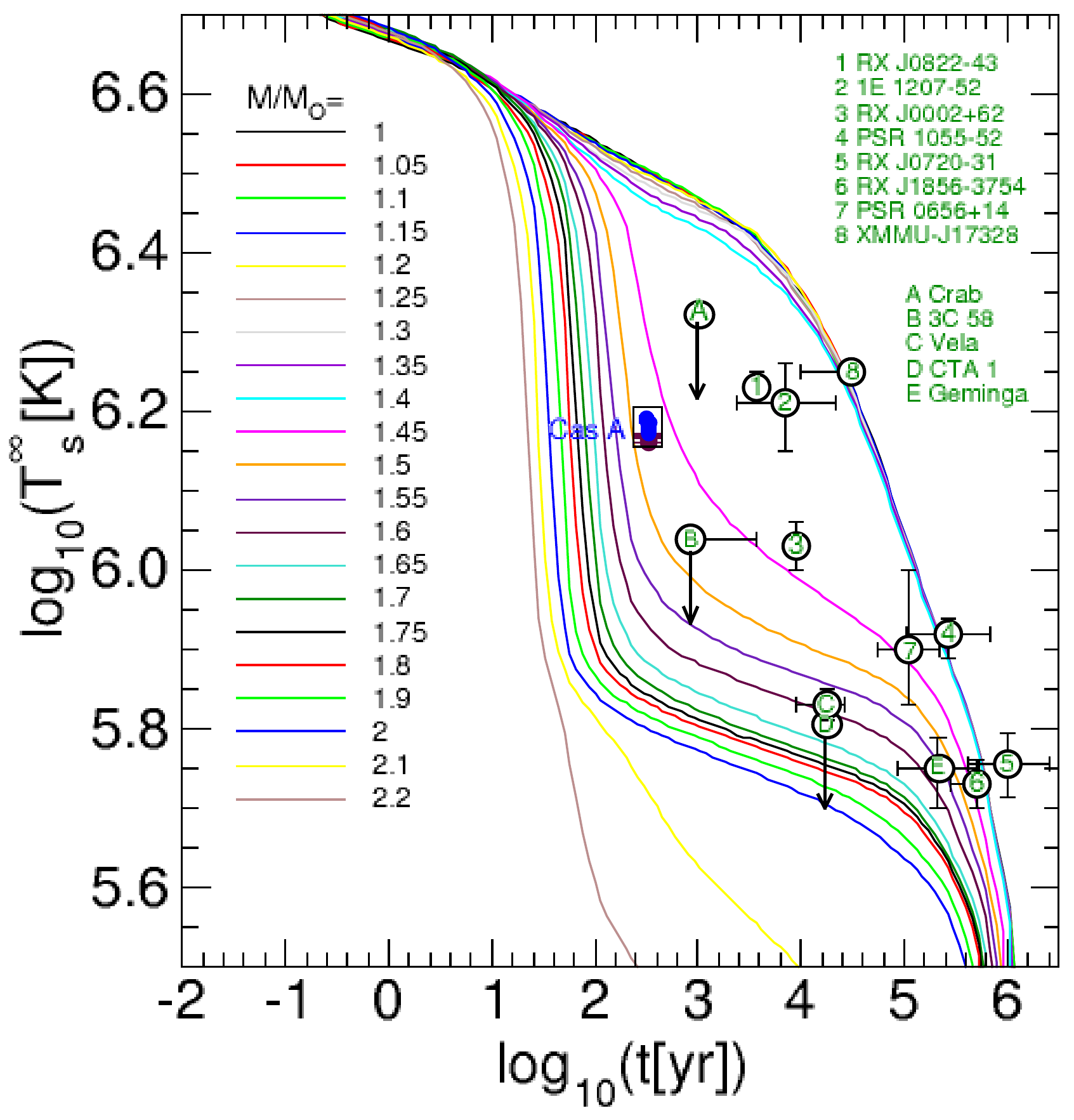}}
\caption{ Redshifted surface temperature as a function of the neutron star age for various neutron star masses and choice of the EoS. Left panel: MKVOR model without the inclusion of hyperons. Right panel: MKVORH$\phi$ model with hyperons included with the gaps following from the TN-FGA parameter choice. Proton gaps for both calculations without and with hyperons are taken following EEHO model.}
\label{F7nohyperon}
\end{figure}

On the left panel in Figure \ref{F7nohyperon} we show the cooling history of neutron stars calculated using the EoS of MKVOR model without inclusion of hyperons. The demonstrated calculations employ the proton gap following the EEHO model shown in Figure \ref{Protongaps}, and the solid curve in Figure \ref{piongap} was used for the effective pion gap.
Hyperons are taken following the TN-FGA parameter choice.
With the pion gaps given by the solid and dashed curves and with proton gaps following the EEHO, EEHOR, CCDK,  CCYms, and T curves we also appropriately describe the cooling history of neutron stars within our scenario.

\section{Conclusion}

Thus we have demonstrated that the presently known cooling data  can be
appropriately described within our nuclear medium cooling scenario, under
the assumption that different sources have different masses.

\acknowledgments{The research  was supported    by the Ministry of Education and Science of the Russian Federation within the state assignment,  project No 3.6062.2017/BY.
The work was also supported by Slovak grant VEGA-1/0469/15.
We acknowledge as well the support of the Russian Science Foundation, project No 17-12-01427.}


\end{document}